\documentclass[preprint,preprintnumbers,amsmath,amssymb]{revtex4}

\usepackage{graphicx}% Include figure files
\usepackage{dcolumn}% Align table columns on decimal point
\usepackage{bm}% bold math

\begin{document}

\title[]{Phase transition and exponential growth of the local polarization in
Ca$_{x}$Ba$_{1-x}$Nb$_{2}$O$_6$ ferroelectric}% Force line breaks with \\

\author{Desheng Fu$^{1,2,3}$ and Kensuke Fukusawa$^{1}$}
\affiliation{$^1$Department of Electronics and Materials Science, Faculty of
Engineering, Shizuoka University, 3-5-1 Johoku, Naka-ku, Hamamatsu
432-8561, Japan.\\
$^2$Department of Engineering, Graduate School of Integrated Science
and Technology, Shizuoka University, 3-5-1 Johoku, Hamamatsu
432-8561,
Japan.\\
$^3$Department of Optoelectronics and Nanostructure Science,
Graduate School of Science and Technology, 3-5-1 Johoku, Naka-ku,
Hamamatsu 432-8011, Japan.}%Lines break automatically or can be forced with \\
\email{fu.tokusho@shizuoka.ac.jp}
\homepage{https://wwp.shizuoka.ac.jp/desheng-fu/}

%\author{Mitsuru Itoh}%
%\altaffiliation[Also at ]{%
%Materials and Structures Lab, Tokyo Institute of Technology}%
%\author{Shin-ya Koshihara}
%\affiliation{Department of Chemistry and Materials Science, Tokyo Institute of Technology, Meguro, Tokyo 152-8551,Japan\\
%This line break forced% with \\}%


\begin{abstract}
Ca$_{x}$Ba$_{1-x}$Nb$_{2}$O$_6$ (CBN) ferroelectric  is isostructural with Sr$_{x}$Ba$_{1-x}$Nb$_{2}$O$_6$ (SBN)  that is  a prototypical ferroelectric of tetragonal  tungsten
bronze (TTB) structure,   but shows a much higher Curie temperature $T_{c}$  than  SBN, providing  a wider operation  temperature for electro-optical  devices.  Here, we report the phase diagram and phase transition of CBN solid solution that are still lack of understanding. We show that CBN  can be  stably formed within a solid solution range of $0.19\leq x\leq0.32$. In sharp contrast to SBN that shows crossover from ferroelectric to  relaxor around $x \approx 0.5$,  CBN merely exhibits a well-defined  first-order phase transition in the whole solid solution range but shows 
polarization precursor dynamics before transition into the ferroelectric phase.
We found that the local polarization  exponentially grows   with temperature  from the Burns temperature $T_{ \rm B}$ toward 
$T_{\rm c}$.  The local polarization was estimated to grow up to several tens pC/cm$^2$ before transition into the ferroelectric phase.  Moreover, we  found that the substitution of Ca for Ba lowers  $T_c$ and  $T_{\rm B}$ and there is a linear correlation between  $T_c$ and the length of polar $c-$axis, which  can be  attributed to the chemical pressure effects. These 
findings may provide a deep  insight into the nature of 
phase transition in the TTB structure oxides.

\end{abstract}

%\pacs{77.65.-j,77.65.Bn, 77.65.Fs, 77.84.Cg, 77.80.Dj}% PACS, the Physics and Astronomy
\maketitle

\section{Introduciton}

\begin{figure}
\includegraphics[width=10cm]{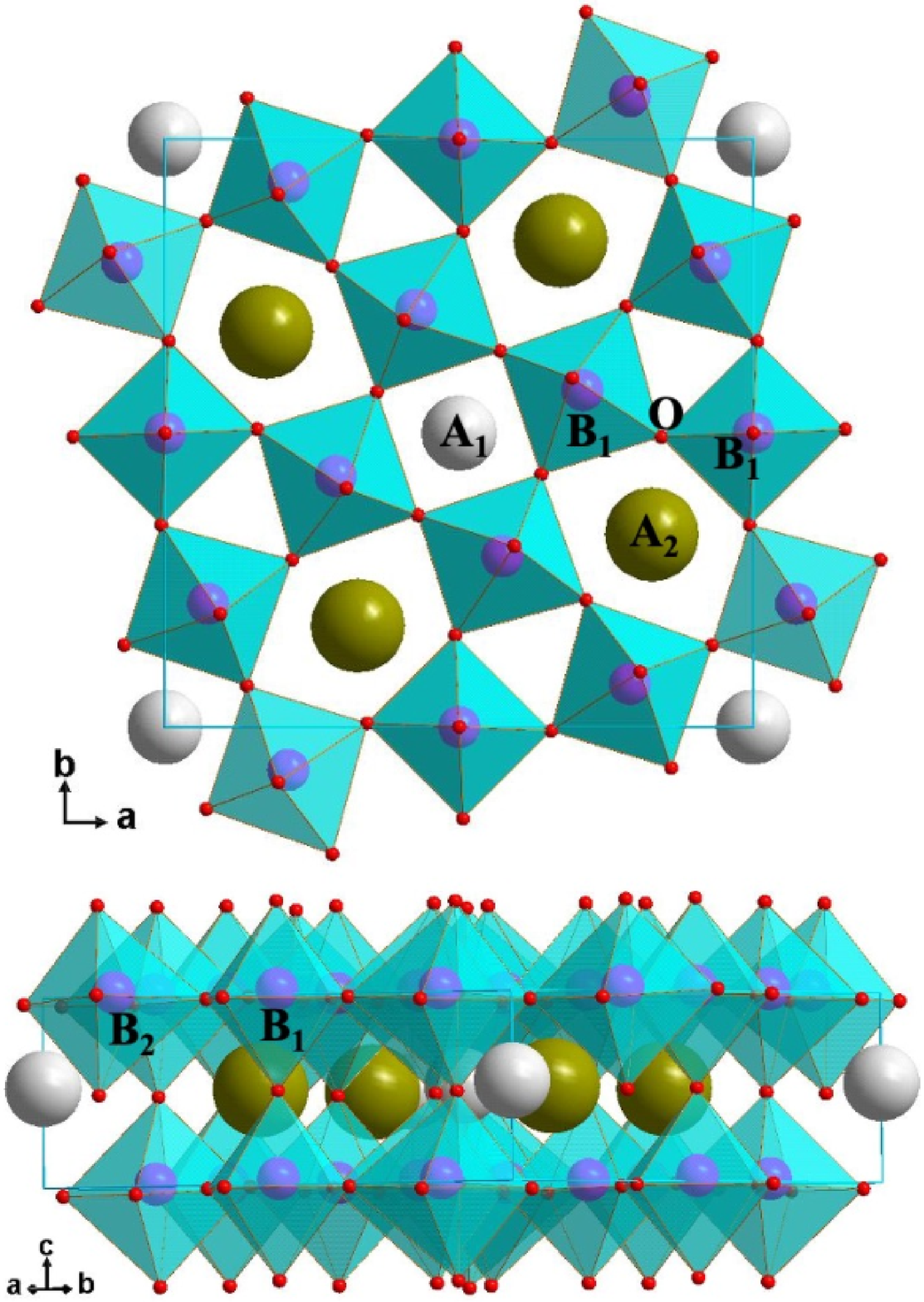}% Here is how to import EPS art
\caption{\label{Fig1} (A$_1)_2$(A$_2)_4$(B$_1)_2$(B$_2)_8$O$_{30}$
 tetragonal  tungsten
bronze  structure.  For Ca$_x$Ba$_{1-x}$Nb$_2$O$_6$ ($0.19\leq x\leq0.32$),  A$_1$, A$_2$ and B$_1$/B$_2$ sites are
occupied by Ca, Ba and Nb, respectively.}
\end{figure}

Ferroelectric with the  (A$_1$)$_2$(A$_2$)$_4$(B$_1$)$_2$(B$_2$)$_8$O$_{30}$
tetragonal tungsten bronze (TTB) structure as shown in Fig.\ref{Fig1} is a
large class of ferroelectric materials and the larger  freedom of
A$_1$, A$_2$, B$_1$ and B$_2$ site element selection allows  one  to
easily  search for the novel ferroelectric in this class of  materials 
as compared with the ABO$_3$ perovskite oxides. Srontium
barium niobate  Sr$_{x}$Ba$_{1-x}$Nb$_{2}$O$_6$ (SBN, $0.2\leq
x\leq 0.8$) \cite{Carrthers} is a prototypical TTB structure ferroelectric
and has attracted much attention due to its excellent pyroelectric,
dielectric, electro-optical and non-linear optical
properties. More interestingly,  SBN shows a crossover from normal
ferroelectric to relaxor around  $x \sim 0.5$, below which  SBN
exhibits typical relaxor behaviors and a well-defined
paraelectric-ferroelectric transition disappears.\cite{Glass,Kleemann1,Kleemann2,Blinc,Banys}. Such relaxor
behavior is believed to be driven by the  intrinsic charge disorder
in the TTB structure. In SBN,  the 12-fold correlated  A$_1$ site is
only  occupied by Sr that has a  smaller ionic radius than Ba, on the other hand the 15-fold correlated A$_2$ site is randomly occupied  by both Sr and Ba.\cite{Jamieson,Podlozhenov,Chernaya1,Chernaya2}  As comparing the chemical
formula  of SBN with the standard one  mentioned above, it  can  be seen that 
1/6 of A$_1$ and A$_2$ -sites are empty in SBN. Such unfilled
tungsten bronze structure is considered to create the charge
disorder, and finally the random field in the crystal,  leading to 
breakdown of  the long-range ordering of spontaneous polarization and the occurrence  of relaxor behaviors in the  Sr-rich SBN. But one
critical issue remains to be addressed   for this scenario : why does
the Ba-rich SBN  show the  well-defined paraelectric-ferroelectric
phase transition rather than the relaxor behavior?.\cite{Glass}  On
the basis of structure analysis, it has been proposed that  relaxor
behaviors  in  SBN  are primarily determined by  the
occupation ratio  of Ba and Sr in the A$_2$ site and the high  ratio of 
Ba occupancy in A$_2$ site will suppresses  the occurrence  of relaxor
behaviors.\cite{Chernaya2}

 Ca$_{x}$Ba$_{1-x}$Nb$_{2}$O$_6$(CBN) is isostructural with SBN and has been  received considerable  attention owing to its excellent electro-optic properties and  a  higher  Curies temperature  than
SBN.\cite{Ebser,Song,Burianek,Muehlberg,Qi,Niemer,Ebser2,Heine,Sheng,Pandey,Suzuki,Pandey2,Graetsch1,Graetsch2} Most of  investigations on CBN  are focused on the crystal growth, dielectric, ferroelectric, optic and elastic properties  of CBN single crystal with congruent melting composition ($x$=0.28).\cite{Ebser,Song,Qi,Niemer,Ebser2,Heine,Sheng,Pandey,Suzuki} In CBN, structural analysis indicates that the A$_1$ site is  exclusively occupied by Ca  and the A$_2$  site predominately by Ba and  the shift of Nb within the NbO$_6$ octahedron along the $c$-axis is responsible for the spontaneous polarization.\cite{Graetsch1,Graetsch2}  Although several reports argue that CBN is relaxor on the basis of the deviation  of  the linear temperature dependence of lattice strain
and the anomalous thermal expansion in the crystal.\cite{Pandey,Suzuki,Pandey2}  It seems that relaxor behaviors are not expected to occur in CBN owing to the predominate A$_2$-site occupancy of Ba with larger ionic radius.\cite{Graetsch1,Ebser2}  However, this issue  remains to be addressed carefully and is the focus of the work.

In this study,  we have  investigated the
phase transition of CBN solid solutions by the dielectric measurements and
the differential scanning calorimetry (DSC). We  clearly demonstrated
that CBN has a well-defined first-order ferroelectric transition within its solid solution range.  This result  is completely different from the  proposed relaxor picture for CBN\cite{Pandey,Suzuki,Pandey2} because no well-defined macroscopic  phase transition should be expected in relaxor.  We further demonstrated that the polarization precursor exists  over a large temperature range above $T_{\rm c}$  and the 
local polarizations  exponentially grow with temperature before transition into the ferroelectric phase. These  findings are of benefit to the  understanding of the phase transition behaviors  in  TTB ferroelectrics.

\section{Experimental}

CBN ceramic samples were prepared  from a  solid-state reaction method. To determine the solid solution range,  we have investigated the phase formation
within a large composition range of $x=0.15-0.40$. The ceramics were sintered  at 1693 K (approximately 50 K lower than the melting point) for 3 hours  to form   dense  samples for the dielectric measurements. Crystal size in the ceramics was estimated to
be  10$\mu$m $\sim$ $50\mu$m from the  Scanning Electron
Microscopy measurements.  Phase formation was
determined by the powder X-ray diffractions(XRD), from which the lattice parameters were calculated with  Si calibration. The  dielectric measurements  were
carried out to observe the phase transition in CBN within a temperature range of 300 K - 870 K by using Agilent 4980A LCR meter. Differential Scanning Calorimetry (DSC) measurements were also performed to  confirm the phase transition by using SII DSC6220 differential scanning calorimetry.

\section{Results and Discussions}

\subsection{Phase formation}

\begin{figure}
\includegraphics[width=16cm]{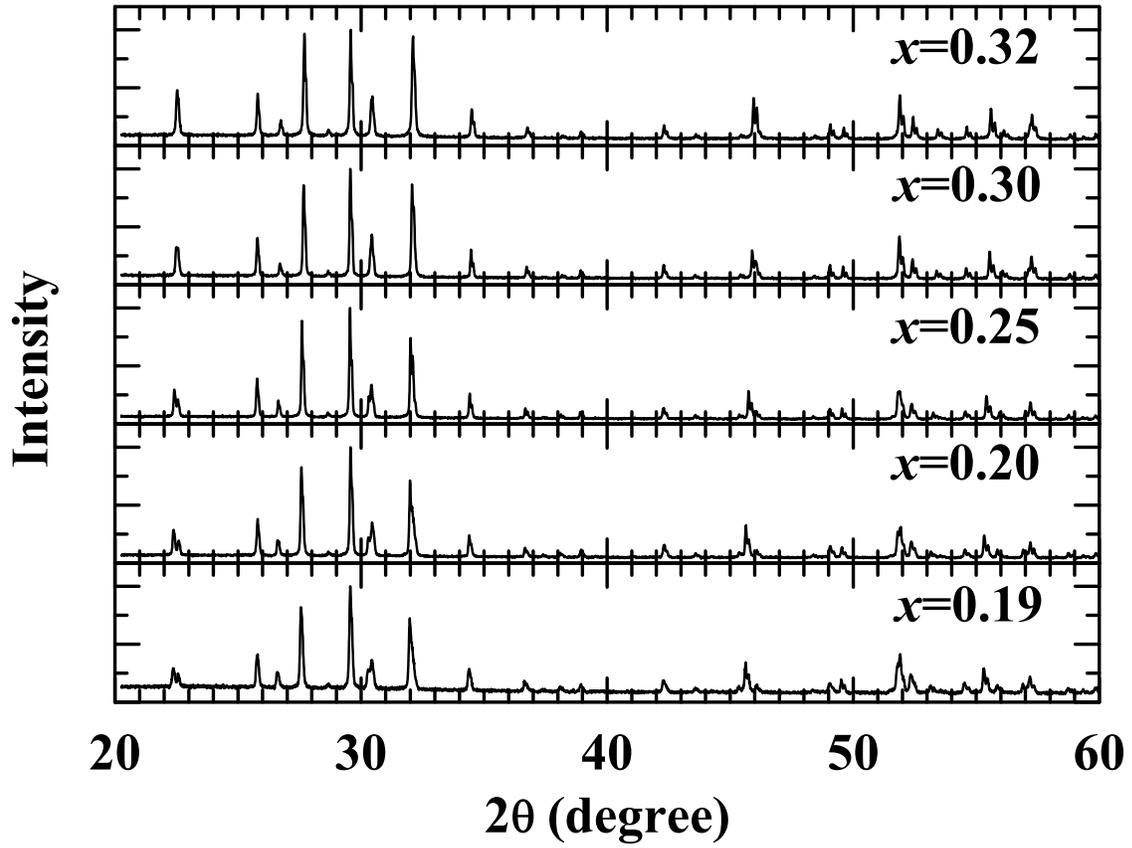}% Here is how to import EPS art
\caption{\label{fig2}  Powder X-ray diffraction patterns of
Ca$_x$Ba$_{1-x}$Nb$_2$O$_6$ within the solid solution range of  $0.19\leq x \leq
0.32$ beyond which mixed phases were found to  coexist  in the compound.}
\end{figure}

To determine the  solid solution range of CBN ferroelectric oxides,
we have examined the phase formation within a  composition range of 
 $0.15\leq x \leq 0.40$. We found that  CBN with pure TTB phase  was only
stably formed   within  the  composition range of $0.19\leq x\leq0.32$ as shown in the XRD patterns in Fig.\ref{fig2},
Out of this range, CaNb$_2$O$_6$- or BaNb$_2$O$_6$-type non-ferroelectric  phase with
orthorhombic structure was  found to coexist with
the CBN ferroelectric phase. The solid solution range of CBN is much
narrower than  that of SBN ( $0.2\leq
x\leq 0.8$) \cite{Carrthers} , which can be attributed to the smaller ionic
radius of Ca (1.34  {\AA}) than that of Sr  (1.44 {\AA}).\cite{Shannon}  This is similar to the  case of the solid
solution between  BaTiO$_3$ and SrTiO$_3$/CaTiO$_3$, in which Ba can
be completely substituted with Sr but only partially with
Ca.\cite{Basmajian,Fu1,Fu2}

\begin{figure}
\includegraphics[width=16cm]{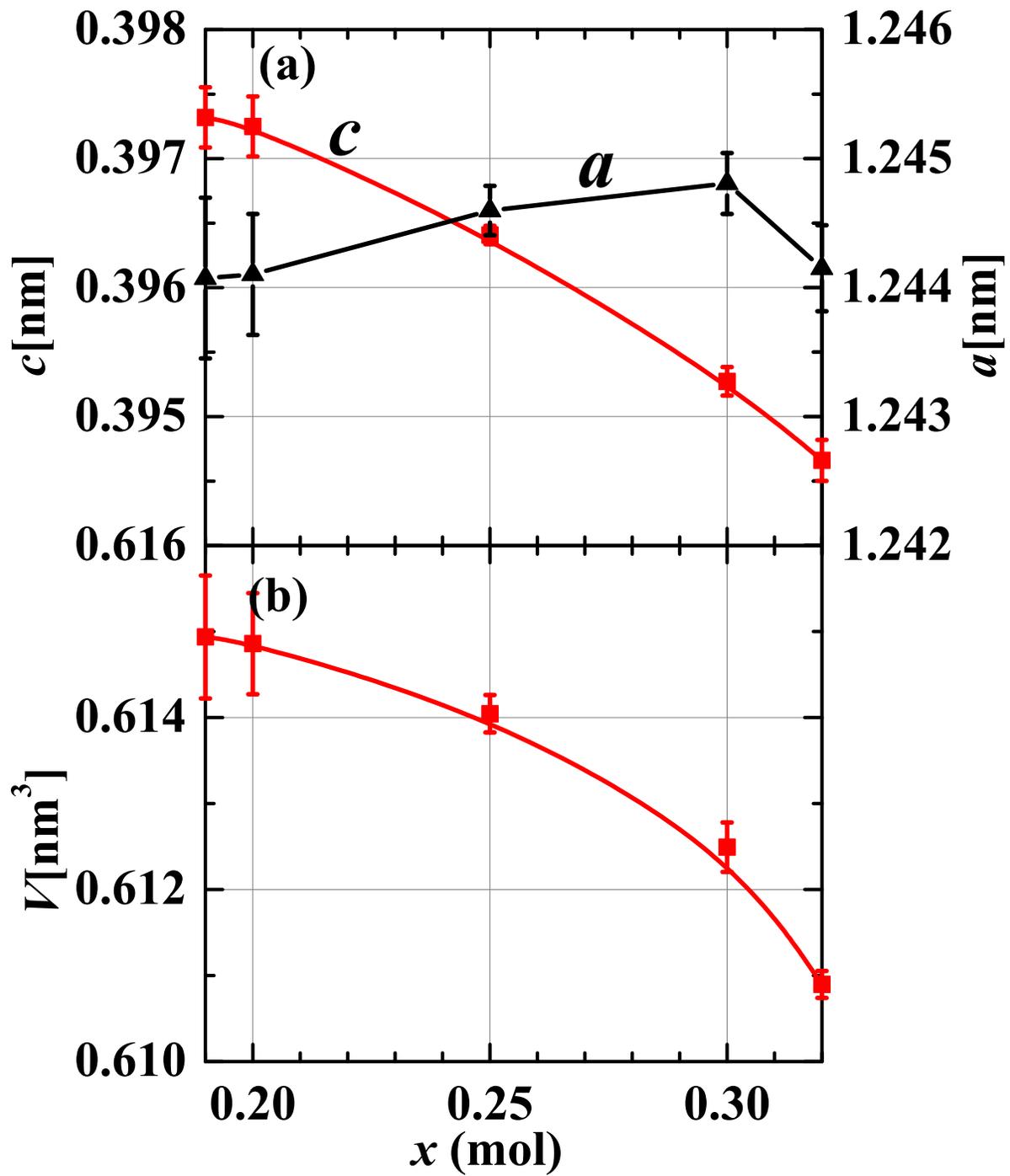}% Here is how to import EPS art
\caption{\label{fig3} Change of the  lattice parameters of
Ca$_x$Ba$_{1-x}$Nb$_2$O$_6$ with composition.}
\end{figure}

In CBN, $a$-axis lattice remains  nearly unchanged within  the solid solution range as shown in Fig.\ref{fig3}.  On the other hand, the polar $c$-axis
lattice is shortened  as increasing the Ca concentration, resulting in
a decrease of the unit cell volume. Because Ca has a smaller ionic radius (1.34  {\AA}) than Ba (1.61  {\AA}),\cite{Shannon} the substitution of Ca for Ba will result in a chemical pressure, leading to the reduction of the unit cell volume. 
The  unit cell volume is suppressed by approximately 0.65\% within the solid solution range, corresponding  to an
average suppression ratio of  5\%/mol. 
These changes owing to chemical pressure  are  much different from the case of Ca substitution for Ba in Ba$_{1-x}$Ca$_x$TiO$_3$ (BCT)  perovskite oxides, in
which both the $a$- and  $c$-axis
lattices of the tetragonal structure show the  same reduction ratio  with
the Ca substitution and the  unit cell volume is reduced by a ratio
of  11\%/mol.\cite{Fu1,Fu2} Therefore,  the chemical pressure in BCT pervskite ferroelectric  is of three
dimension, in contrast, is of one-dimension  in CBN TTB ferroelectric.  Since both the  frameworks of perovskite
and TTB structures are  formed by the BO$_6$ oxygen octahedron and TTB
structure can be considered to be derived from the A-site vacancy of
the perovskite structure. It seems that  the different chemical pressure effects
between TTB CBN and  perovskite BCT may be related to the 1/6 A-site
vacancy in the  TTB structure.

\subsection{Phase transition}

\begin{figure}
\includegraphics[width=16cm]{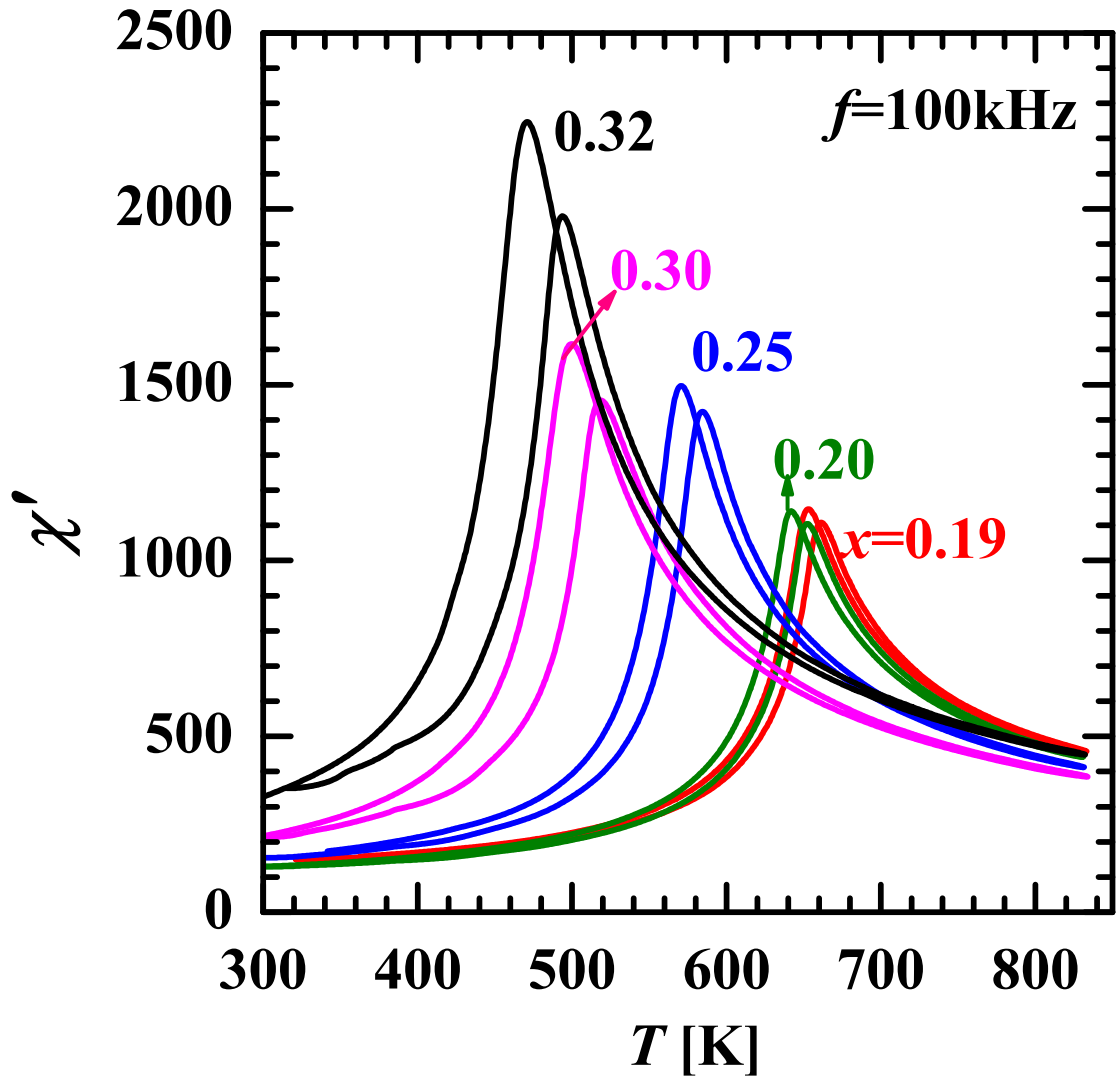}% Here is how to import EPS art
\caption{\label{fig4} The dielectric susceptibilities of
Ca$_x$Ba$_{1-x}$Nb$_2$O$_6$ ceramics as a function of
temperature and composition.}
\end{figure}

\begin{figure}
\includegraphics[width=16cm]{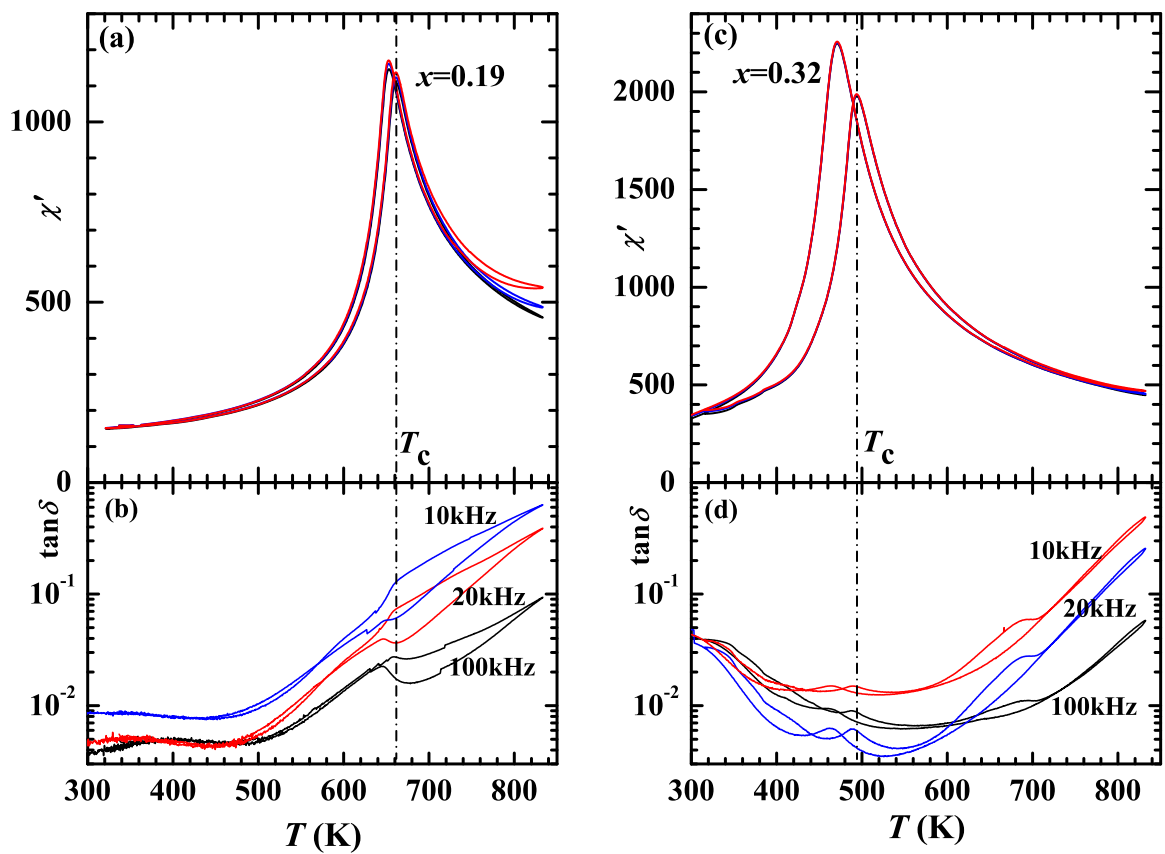}% Here is how to import EPS art
\caption{\label{fig5}   Temperature variation of dielectric
susceptibility and loss  for different frequencies in Ca$_x$Ba$_{1-x}$Nb$_2$O$_6$. It is clear that the transition temperature  does not depend on frequency and has a thermal hysteresis.}
\end{figure}

To investigate the phase transition in CBN, we have investigated
the variation of dielectric susceptibility for different frequencies
as a function of temperature within a temperature range of $300
\sim840$ K and the results were shown in Fig.\ref{fig4} and
Fig.\ref{fig5}, respectively. Within the solid solution range, CBN
shows a  well-defined and frequency-independent  phase transition.
This is essentially different from  relaxor such as
 the prototypical relaxor Pb(Mg$_{1/3}$Nb$_{2/3}$)O$_3$
(PMN), which shows a broad peak of dielectric susceptibility with
strong frequency dispersion in the radio frequencies over a large
temperature range \cite{Fu3,Fu4,Miga,Levstik} and a  smear
ferroelectric phase transition\cite{Fu3,Fu4,Taniguchi} without any
detectable heat capacity peak.\cite{Moriya}

Fig.\ref{fig4} and Fig.\ref{fig5} also clearly indicates that the
ferroelectric phase transition in CBN is a  first-order phase
transition that  has large thermal hysteresis.
For the cooling/ heating rate of 2 K/min used  in the measurements,  the
difference of $T_{\rm c}$ between heating and cooling is 12.4 K for
$x=0.19$ and  increases to 25.2 K for $x=0.32$. This indicates
that the substitution of Ca for Ba in CBN enhances the thermal
hysteresis of the phase transition.

\begin{figure}
\includegraphics[width=16cm]{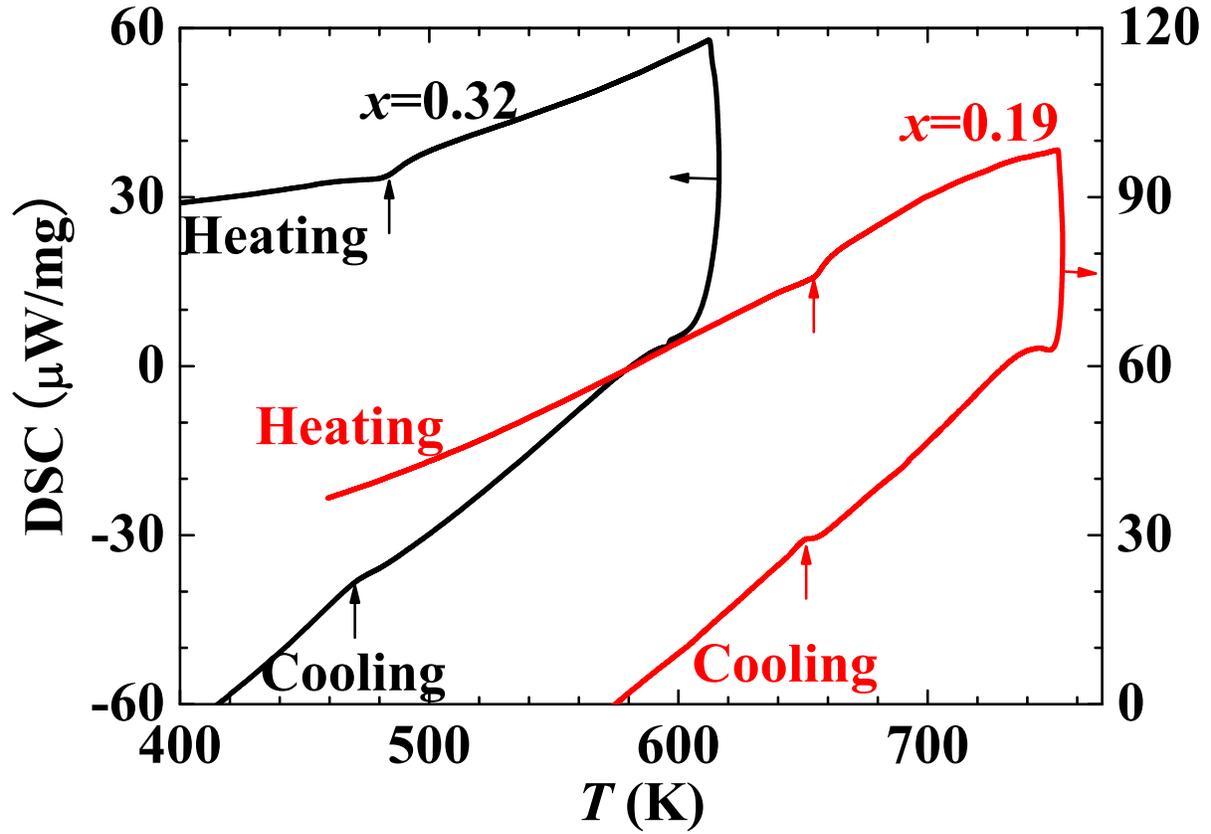}% Here is how to import EPS art
\caption{\label{fig6}  Phase transition in
Ca$_x$Ba$_{1-x}$Nb$_2$O$_6$ detected by differential scanning
calorimetry (DSC). Exotherm or endotherm was seen during the phase
transition as indicated by the arrows. }
\end{figure}

To further confirm the nature of thermal phase transition in CBN,
we also observed the change in enthalpy during the phase transition
by the DSC measurements. The results are shown in Fig.\ref{fig6}.
Indeed, we observed a clear change in enthalpy during the phase
transition. DSC measurements also  confirm  that this phase
transition is  of first order.  DSC measurements results
are in good agreements with those of the dielectric
measurements, and also consistent with the heat capacity
measurements  reported for CBN crystal with $x=0.31$, in which a
sharp peak of heat capacity has been clearly observed during the phase transition.\cite{Muehlberg} These results are also completely different from that observed for relaxor such as the 
prototypical relaxor PMN,
in which no a  peak of heat capacity can be  probed  around the  peak temperature of the dielectric response. \cite{Moriya}

All  the above facts  clearly indicate that CBN crystal undergoes a well-defined  thermal  phase transition that is of first-order and has a large  thermal
hysteresis of approximately  $12.4 \sim25.2$ K.

\subsection{Polarization precursor behaviors}

Polarization precursor is commonly   observed in normal ferroelectric such as BaTiO$_3$  before transition from paraelectric phase to  ferroelectric phase.\cite{Burns,Takagi,Tai,Ziebinska,Namikawa,Dulkin,Ko,Pugachev}  Theoretically, it has been proposed that local dynamical
precursor domains commonly occur in the paraelectric phase of the perovskite ferroelectric.\cite{Bussmann-Holder1,Bussmann-Holder2}  Similar to the occurrence of polar nano-regions (PNRs)  in relaxor at the so-called Burns's temperature    $T_{\rm B}$,
polar domains occur in paraelectric phase  from $ T_{\rm B}$ and  grow upon approaching $T_{\rm c}$ to form a
homogeneously polarized state below $T_{\rm
c}$. In contrast to PNRs of relaxor that cause the very strong frequency
dispersion of dielectric susceptibility,
polarization precursor domains in the perovskite ferroelectric such as BaTiO$_3$
do not  give rise to a frequency-dependent dielectric
response in the radio frequency.\cite{Bussmann-Holder1}

For CBN single crystal with congruent
melting composition $x$=0.28, the birefringence has been
demonstrated to occur in a temperature region of  $T_{\rm
c}<T<T_{\rm c}+ 40 $K.\cite{Ebser2} The birefringence in the paraelectric phase  is commonly attributed to the polarization fluctuation $\langle P_c^2\rangle- \langle
P_a^2\rangle$ in the crystal\cite{Lehnen,Takagi}
 \begin{equation}\label{eq1}
\Delta n=-(n^3_0/2)( g_{11}-g_{12})\langle P^2 \rangle,
 \end{equation}
where $g_{11}$ and  $g_{12}$  are the electro-optic coefficients. In addition to the occurrence of birefringence, the
deviation from the linear temperature dependence of lattice strain
and the anomalous thermal expansion also  have been reported for CBN with $x=0.28$.\cite{Pandey} These results indicate that polarization precursor indeed  exists    in the paraelectric phase of CBN
crystal with $x=0.28$.

\begin{figure}
\includegraphics[height=18cm]{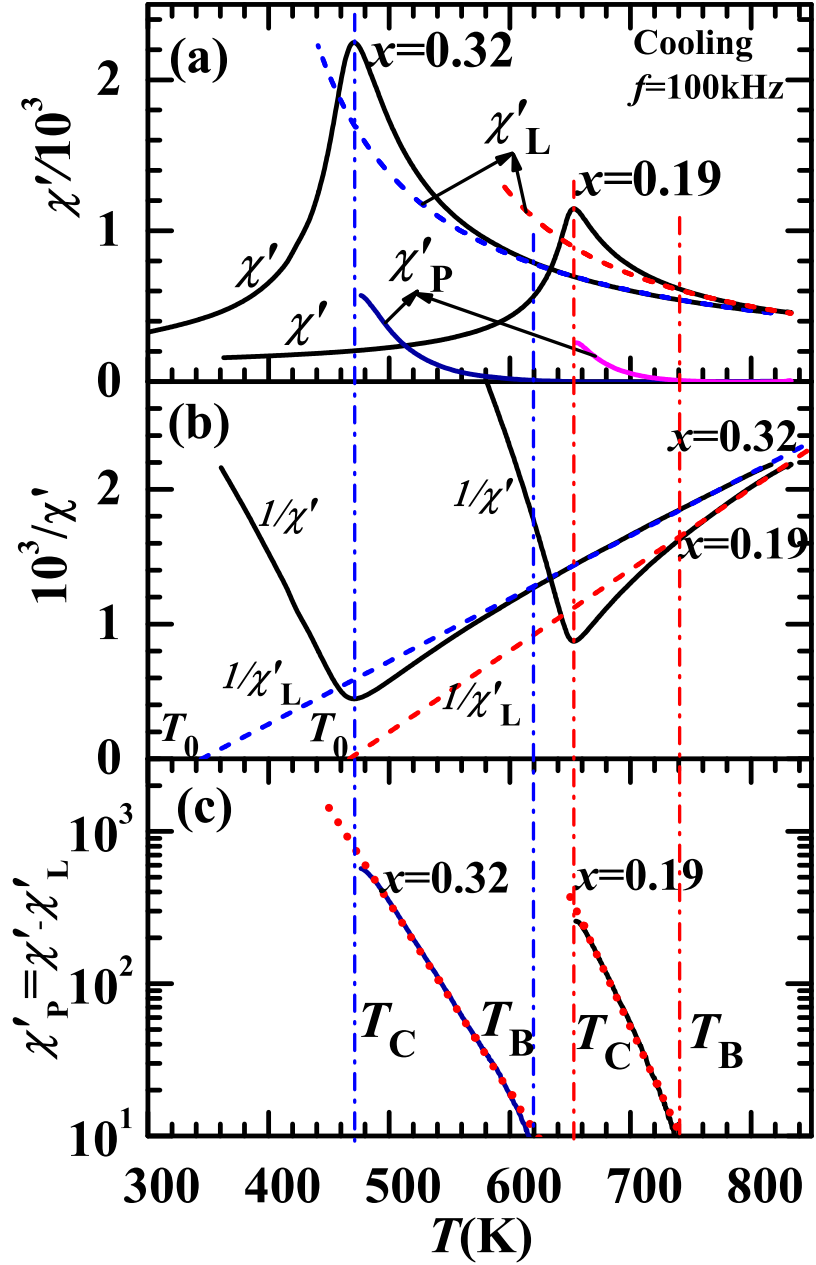}% Here is how to import EPS art
\caption{\label{fig7}  (a)Temperature variations of
the measured dielectric susceptibility $\chi'$, the dielectric susceptibility $\chi'_{\rm L}$  due to the lattice response estimated from the  Curie law (dashed lines), and  the dielectric susceptibility $\chi'_{\rm P}$( $=\chi'-\chi'_{\rm L}$)  due to the local polarization response. (b) Inverse of the
dielectric susceptibility $\chi'$. The fitting results  by the Curie law
are also shown by the dashed lines. Deviation from Curie law was
seen from $T\approx T_{\rm B}$ upon cooling. (c) Change of $
\chi'_{\rm P}=\chi'-\chi'_{\rm Curie}$ plotted in a logarithmic scale for
$T_{\rm c}<T<T_{\rm B}$. Dotted lines are the fitting results of an
exponential law given in equation (\ref{eq6}).}
\end{figure}

To get more insight into the polarization precursor dynamics
in CBN, we thus performed a  detailed analysis on the
temperature dependence of dielectric susceptibility of CBN
within the whole  solid solution range. To obtain a reliable result, here we  used the  dielectric susceptibility  at 100 kHz that shows a  low dielectric loss up to high temperature of $\sim$800 K  for the analysis.
Upon cooling from $T\approx$ 800 K,   the dielectric
susceptibility $\chi'$ initially obeys the Curie law
\begin{equation}\label{eq2}
    \chi'(T)=C/(T-T_0)\;\; {\rm for}\;\;  T>T_{\rm B},
\end{equation}
where $C$ is the Curie constant and $T_0$ is the Curie-Weiss temperature.
However, as shown in Fig.\ref{fig7}, the  dielectric
susceptibility deviates from  the  Curie law around  the Burns temperature $T_{\rm B}$. The value of $T_{\rm B}$ was estimated to be
approximately $T_{\rm c}$+88 K for $x=0.19$ and $T_{\rm c}$+143 K
for $x=0.32$, respectively.  $T_{\rm B}$ shows a linear
relationship  with $x$ and is decreased as
increasing the Ca-concentration  like   $T_{\rm c}$ (Fig.\ref{fig10}(a)).

\begin{figure}
\includegraphics[height=12cm]{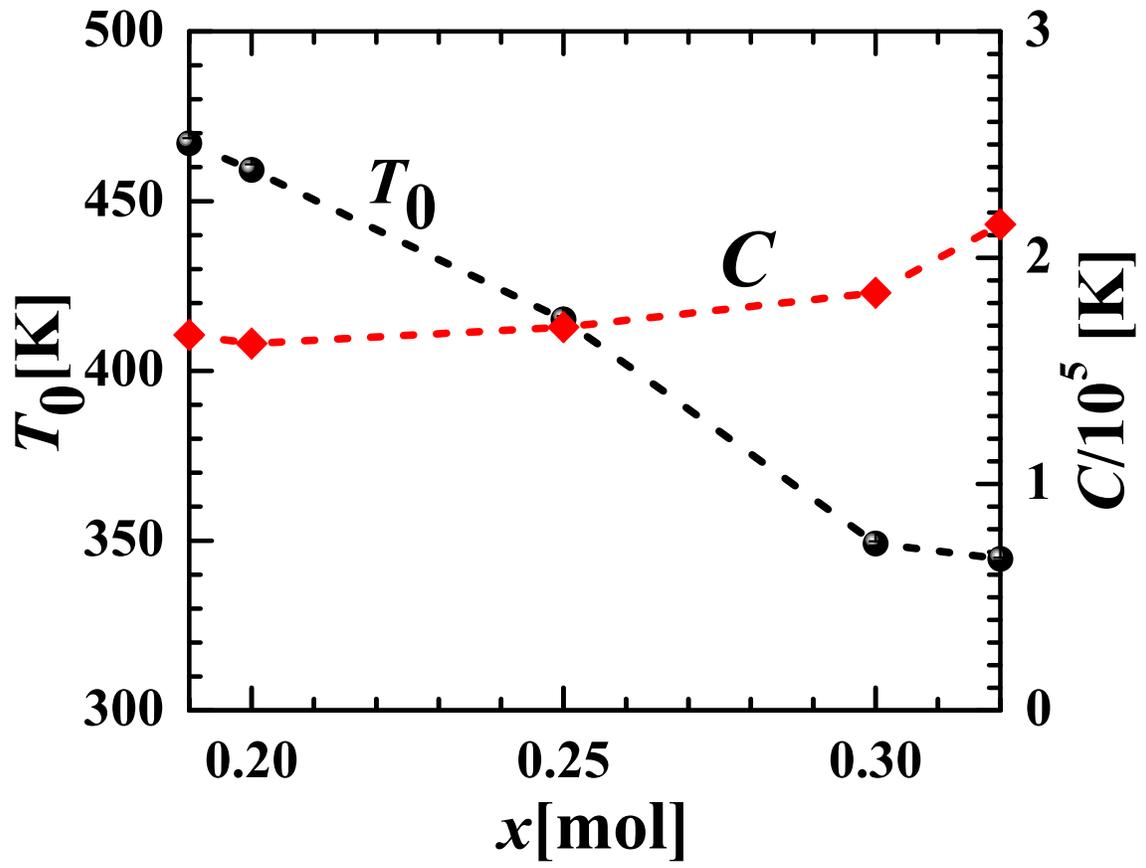}% Here is how to import EPS art
\caption{\label{fig8} The fitting  values of the Curie constant $C$ and the Curie-Weiss temperature $T_0$.}
\end{figure}

The fitting parameters of the Curie law are summarized in
Fig.\ref{fig8}. The  Curie constant shows a slight change with the Ca-concentration and has a value comparable  to that of
BaTiO$_3$ single crystal ($1.5 \times10^5$ K).\cite{Shiozaki,Lines}  Curie constant of the 
displacive type ferroelectrics commonly has a magnitude of 10$^5$ K
that is one order larger than the order-disorder type ferroelectric.\cite{Lines}  Hence, the value of the  Curie constant suggests that  CBN is a displacive
type ferroelectric. Apparently, this conclusion is consistent with the structure
analysis reported  for CBN($x=0.28$), which demonstrates  that the spontaneous polarization in CBN is owing to  the Nb
displacement within the NbO$_6$ octahedron.\cite{Graetsch1}

For  $T>T_{\rm B}$, the dielectric
susceptibility of  CBN exactly follows the  Curie law,
indicating  that the dielectric response in these high temperatures
is  owing  to the lattice dynamics response. However, upon further
cooling, deviation from the  Curie law  was observed from $
T_{\rm B}$. As mentioned above, this deviation of the
dielectric susceptibility can be reasonably attributed to   the
polarization precursor dynamics occurring before the transition into the ferroelectric phase.   For $T_{\rm c}<T<T_{\rm B}$,
the total dielectric susceptibility $\chi'(T)$  thus can be considered to be contributed by the  lattice response  ($\chi'_{\rm L}(T)$)
and  the polarization precursor responses ($\chi'_{\rm P}(T)$), respectively,
 \begin{equation}\label{eq3}
   \chi'(T)=\chi'_{\rm L}(T)+\chi'_{\rm P}(T)\;\; {\rm for}\;\; T_{\rm c}<T<T_{\rm B}.
 \end{equation}
The contribution
of lattice dynamics to the  dielectric responses  can be estimated from the Curie law using equation (\ref{eq2}),
 \begin{equation}\label{eq4}
   \chi'_{\rm L}(T)=C/(T-T_0)\;\; {\rm for}\;\; T_{\rm c}<T<T_{\rm B}.
 \end{equation}
We thus calculated  the contribution of polarization  precursor dynamics to the
dielectric response from  equation (\ref{eq3}),
\begin{equation}\label{eq5}
  \chi'_{\rm P}(T)=\chi'(T)-\chi'_{\rm L}(T)\;\; {\rm for}\;\; T_{\rm c}<T<T_{\rm B}.
 \end{equation}

\begin{figure}
\includegraphics[height=20cm]{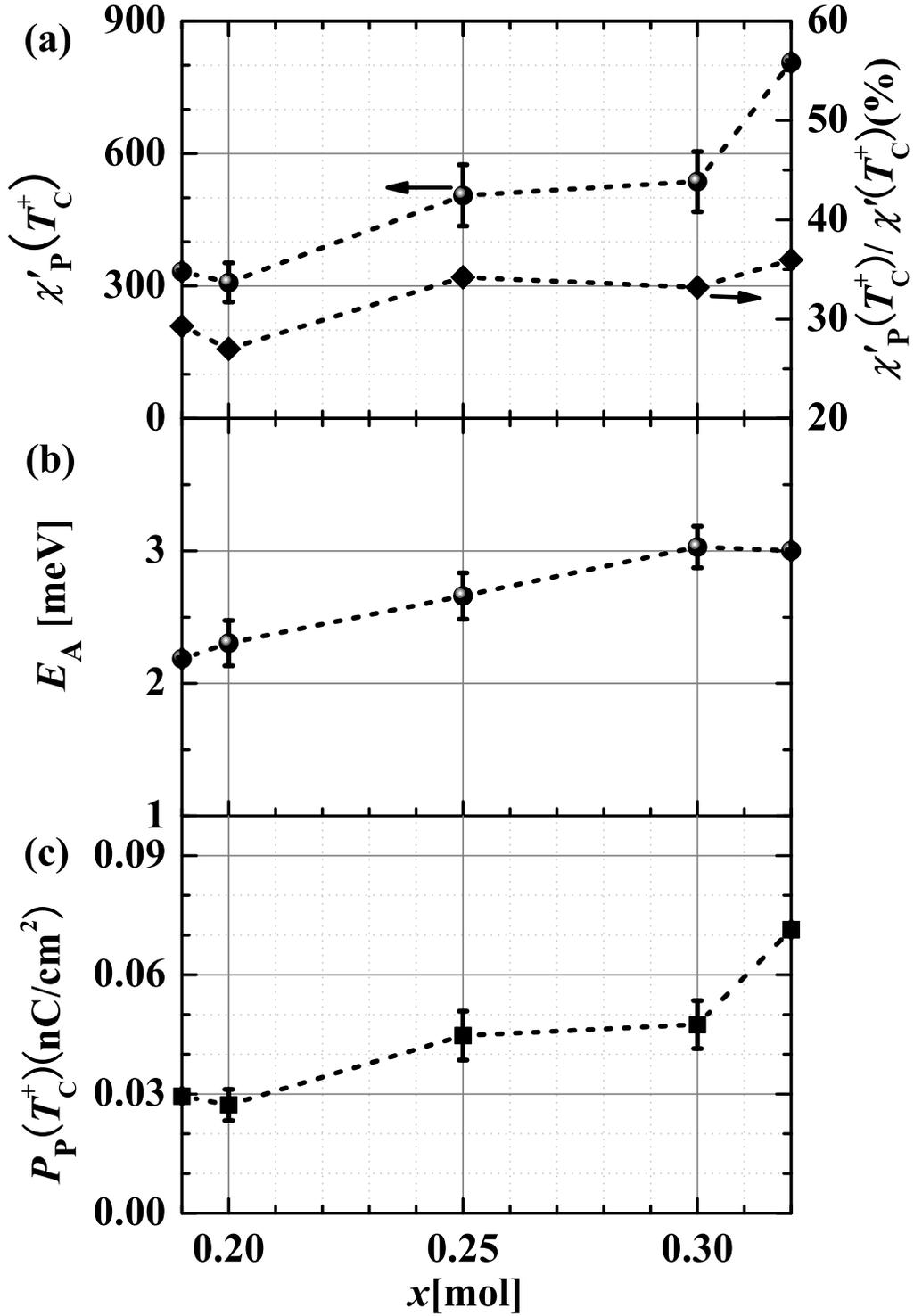}% Here is how to import EPS art
\caption{\label{fig9} (a) The dielectric susceptibility  ($\chi'_{\rm P}(T_c^+)$) due to the local polarization  contribution and its ratio to  the total  dielectric response  at the  transition point 
$T=T_{\rm c}^+$ in the paraelectric phase. (b) Activation energy  $E_{\rm a}$ required for  the
local  polarization  to grow  within the temperature range of   $T_{\rm c}<T<T_{\rm B}$. (c) The estimated local  polarization $P_{\rm P}(T_c^+)$ growing at the transition point 
$T=T_{\rm c}^+$ in the paraelectric phase. }
\end{figure}

\begin{figure}
\includegraphics[height=12cm]{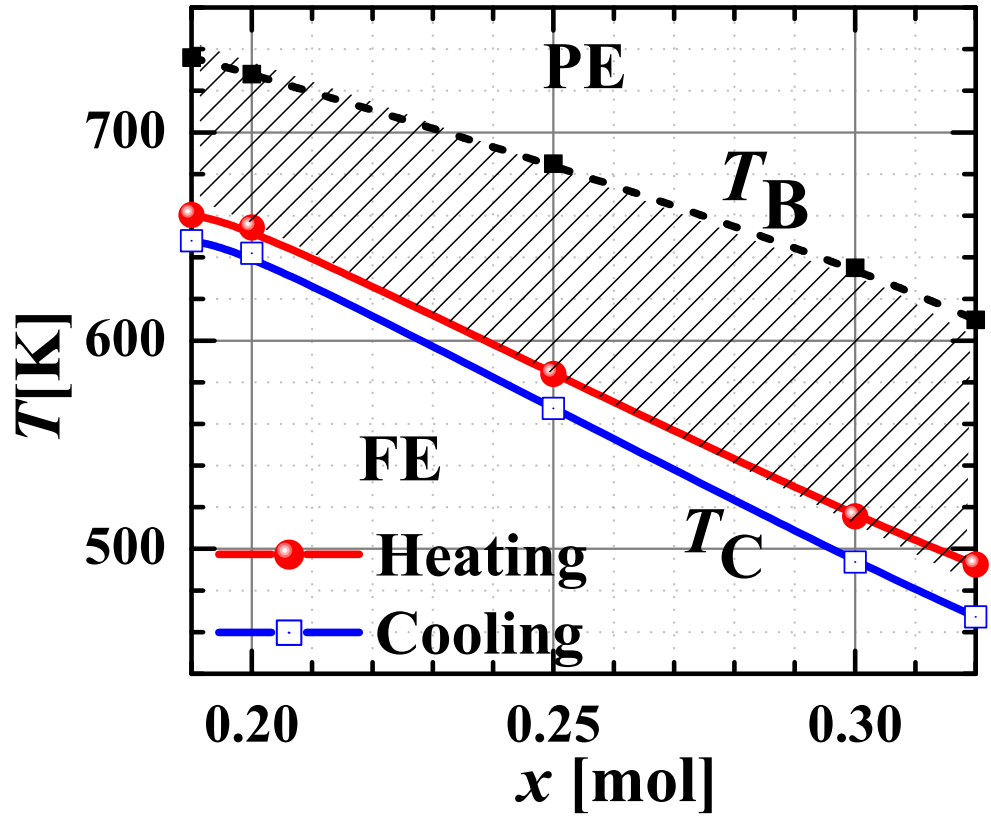}% Here is how to import EPS art
\caption{\label{fig10} A phase diagram proposed for Ca$_x$Ba$_{1-x}$Nb$_2$O$_6$. The shaded region shows  the occurrence  of the  local polarizations  in the paraelectric mother phase. FE=ferroelectric and PE=paraelectric.}
\end{figure}

The temperature variation of $\chi'_{\rm P}(T)$ is shown in
Fig.\ref{fig7}(a) and (c) for the two end compositions of CBN.
$\chi'_{\rm P}$ increases when $T$  is lowered toward the
transition temperature  $T^{+}_{\rm c}$. Surprisingly, we found that $\chi'_{\rm P}(T)$ shows an exponential increase when  $T$ is  lowered toward the transition temperature  $T^+_{\rm c}$. This has been 
clearly shown in Fig.\ref{fig7}(c).    $\chi'_{\rm
P}(T)$ can  be well fitted  by the exponential law for all compositions within the solid solution range,
 \begin{equation}\label{eq6}
    \chi'_{\rm P}(T)=\chi'_{\rm P}({T^{+}_{\rm c})}e^{-k_{\rm B}(T-T_{\rm c})/E_{\rm a}}\;\; {\rm for}\;\; T_{\rm c}<T<T_{\rm B},
 \end{equation}
where $k_{\rm B}$ is Boltzmann constant, $\chi'_{\rm P}(T^{+}_{\rm c})$
and $E_{\rm a}$ are constants.The fitting values of $\chi'_{\rm
P}(T^{+}_{\rm c})$ and $E_a$ were summarized  in Fig.\ref{fig9} (a)
and (b). For  $x=0.19$, it was
estimated that   29 \% of the total dielectric susceptibility at  $T^{+}_{\rm
c}$ was contributed by the polarization precursor. For another end of the solid solution $x=0.32$, the  polarization  precursor contributes approximately 35\% of the total dielectric response at
$T^{+}_{\rm c}$.  It seems that  the Ca-substitution
slightly enhances the contribution ratio  of the polarization precursor dynamics to the
total dielectric response in CBN.

From textbook, we know that  polarization $P$ is
proportional to the dielectric susceptibility $\chi'$,
 \begin{equation}\label{eq7}
    P=\chi'\epsilon_0E,
     \end{equation}
where $E$ is an electric field and $\epsilon_0$ is the dielectric
permittivity  of vacuum.  From equation (\ref{eq6}) and   (\ref{eq7}), we can immediately obtain the relationship of the  locally-grown  polarization $P_{\rm L}(T) $ with the temperature $T$,
\begin{equation}\label{eq8}
   P_{\rm L}(T)= \chi'_{\rm P}(T)\epsilon_0E=\chi'_{\rm P}(T^{+}_{\rm c})\epsilon_0Ee^{-k_{\rm B}(T-T_{\rm c})/E_{\rm a}}\;\; {\rm for}\;\; T_{\rm c}<T<T_{\rm B}.
 \end{equation}
Here, we can clearly see that the local polarization experientially grows when the temperature is lowered toward the phase transition. Now it is  also clear that  $E_{\rm a}$ defined in equation(\ref{eq6})  can be
understood as the activation energy required for the  local
polarization growth in the paraelectric phase.   It was  evaluated
to be 2.2 mev $\sim$ 3 mev for CBN  (Fig.\ref{fig9}(b)).

We then   made  an estimation on the local
polarization growth at $T=T_{\rm c}^+$ by  using $\chi'_{\rm P}({T^{+}_{\rm c})}$ obtained from equation (\ref{eq6}) and  the electric
field ($E=10^{-3}$ kV/cm) used in the dielectric measurements,
\begin{equation}\label{eq9}
  P_{\rm L}(T^{+}_{\rm c})= \chi'_{\rm P}(T^{+}_{\rm c})\epsilon_0E.
 \end{equation}
The results  were given in Fig.\ref{fig9}(c) as a function of
composition $x$. At $T=T_{\rm c}^+$, the local polarization is
predicted to grow to a magnitude of 0.03 nC/cm$^2$ 
for  $x=0.19$  and  0.07 nC/cm$^2$ for $x=0.32$, respectively. As shown in Fig.\ref{fig9}(c), the substitution of Ca for Ba granularly enhances the growth of the local polarization. This   can be reasonably
understood from  the lowering of  $T_{\rm c}$ with the  increase of $x$ because the lowering of $T_c$ allows the  local polarization to grow within  a larger temperature
range ($ T_{\rm c}<T<T_{\rm B}$). This is evident from
Fig.\ref{fig7}(c) and Fig.\ref{fig10}. 

The spontaneous
polarization $P_{\rm S}$  has been reported to be
35.3 $\mu$ C/cm$^2$ for CBN($x=0.28$) at room temperature.\cite{Qi}  On the other hand, the local polarization grown at  $T=T_{\rm c}^+$   has  been estimated to has a value of $0.03-0.07$  nC/cm$^2$ that  corresponds to a  magnitude of  $\sim 10^{-6}\times P_{\rm S}$. This situation is completely
different from the case of the prototypical relaxor PMN. In PMN
single crystal, the polarization has grown  to a magnitude of
$\sim\mu$C/cm$^2$ around the broad peak  of dielectric response (
$T\approx$ 260-270 K)  and remarkable remanent polarization  has
been observed in these temperatures, indicating a large volume fraction of PNRs  in the relaxor crystal at the peak temperature.\cite{Fu3,Fu4,Levstik}

\subsection{Phase diagram}
On the basis of the results on the  phase transition and the  polarization precursor
described above, we then proposed a phase diagram
 for CBN solid solutions and showed it in 
 Fig.\ref{fig10}. For temperatures higher than  $T>T_{\rm B}$,  CBN  can be  considered to be  under a  pure
 paralectrics state. After  cooling to $T\approx T_{\rm
 B}$, polarization precursors emerge in the paraelectric mother
 phase,  these local
 polarizations exponentially grow  as lowering $T$ to  the transition temperature 
 $T_{\rm c}$. The ferroelectric phase transition is of
 first-order, and the transition temperature upon heating is
 $12.4\sim 25.2$ K higher than that upon cooling. As shown in the
 phase diagram,  increase in Ca-concentration leads to the lowering of $T_{\rm c}$ and $T_{\rm B}$,   a larger thermal
 hysteresis and a larger temperature range for the local polarization growth.

In addition, the substitution of Ca for Ba results in a nearly
linear reduction of $T_{\rm c}$ and $T_{B}$ with $x$. As well known
in many normal ferroelectrics, pressure can significantly reduces the
ferroelectric transition temperature.\cite{Fu1,Ishidate} In solid solution, chemical pressure is commonly considered to be produced owing to the chemical substitution of the smaller ions for
the larger one such as the substitution of Ca for Ba in CBN. The chemical pressure generally reduces the unit cell volume, and finally results in the decrease of $T_{\rm c}$ in ferroelectric such as (Ba$_{1-x}$Sr$_x$)TiO$_3$.\cite{Shiozaki}  Therefore, the reduction of $T_{\rm c}$ and $T_{\rm B}$ with the substitute of Ca for Ba in CBN can be reasonably explained by such chemical pressure effects.

\begin{figure}
\includegraphics[height=12cm]{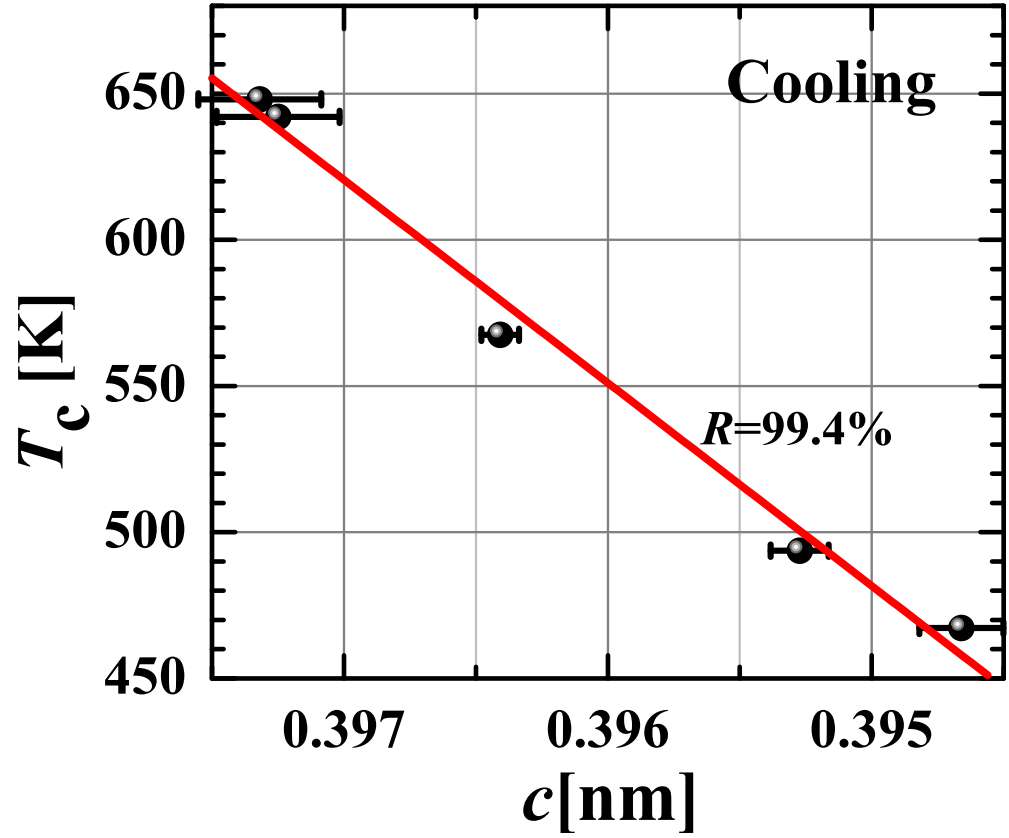}% Here is how to import EPS art
\caption{\label{fig11} The linear correlation between $T_{\rm c}$ and $c$-axis length in  Ca$_x$Ba$_{1-x}$Nb$_2$O$_6$. The correlation degree was estimated to be  99.4\% }
\end{figure}

Interestingly, we found that there is a linear correlation between
$T_{\rm c}$ and the length of polar $c$-axis as shown in
Fig.\ref{fig11}. Structure analysis indicates that the
spontaneous polarization in CBN is originated from the Nb displacement
within the NbO$_6$ octahedron along the $c$-axis in the TTB
structure.\cite{Graetsch1} As  mentioned in section III(a), the
chemical pressure due to the substitution of Ca for Ba leads to the
 compression of $c$-axis, which will reduce the space available for the 
Nb displacement and finally lead to the reduction of
the ferroelectricity and  the $T_{\rm c}$.

\section{Summary}

In summary, we have studied the phase transition in Ca$_x$Ba$_{1-x}$Nb$_{2}$O$_6$(CBN)  TTB ferroelectric
within its solid solution range of   $0.19\leq x\leq0.32$. We demonstrated  that CBN  exhibits a well-defined first-order ferroelectric phase transition, which is in sharp contrast to the  isostructural Sr$_x$Ba$_{1-x}$Nb$_{2}$O$_6$(SBN)  that shows typical relaxor behaviors for Sr-rich composition.  We also show that the local polarization occurs from the Burns temperature $T_{\rm B}$ in the paraelectric  mother phase of CBN. Such local polarization was found to exponentially grow  with temperature before transition into the ferroelectric phase. The local polarization was estimated to grow up to several tens pC/cm$^2$ before transition into the ferroelectric phase.  A phase diagram was then established for
CBN ferroelectric solid solution.  We also found that the substitution of Ca for Ba lowers the transition temperature $T_c$ and Burns temperature $T_{\rm B}$, which has been explained by the suppression of the polar $c$ axis owing to the chemical pressure. Moreover, a linear correlation between $T_c$ and the length of polar $c-$axis was found in CBN. These findings  provide new insights into the  understanding of the underlying physics in the ferroelectric with TTB structure.

\newpage %Just because of unusual number of tables stacked at end

% \bibliography{}% Produces the bibliography via BibTeX.

%\ack{This work was partly supported by Grants-in-Aid for Scientific
%Research from The Ministry of Education, Culture, Sports, Science
%and Technology of Japan}

%\clearpage
%\bibliography{BCZT-piezo}% Produces the bibliography via BibTeX.

\clearpage
%\begin{references}

%\end{references}

\clearpage \clearpage
\newpage

\newpage

\newpage

\newpage

\newpage

\newpage

\newpage

\newpage

\newpage

\newpage

\newpage

\end{document}